\documentclass[9pt,twocolumn,twoside]{osajnl}
\journal{optica} % Choose journal (ao, aop, josaa, josab, ol, optica, pr)

\setboolean{shortarticle}{false}
\RequirePackage{siunitx}
\RequirePackage{xcolor}
\RequirePackage{amsmath}
\DeclareSIUnit\photons{ph}
\DeclareSIUnit\particles{particles}

\title{Ptychographic wavefront characterisation for single-particle imaging at X-ray lasers}

\author[1,2,3]{Benedikt J. Daurer}
\author[3,4,5]{Simone Sala}
\author[6]{Max F. Hantke}
\author[2]{Hemanth K.N. Reddy}
\author[2,7]{Johan Bielecki}
\author[1,8]{Zhou Shen}
\author[2,9]{Carl Nettelblad}
\author[2,10]{Martin Svenda}
\author[2]{Tomas Ekeberg}
\author[11,12]{Gabriella A. Carini}
\author[11]{Philip Hart}
\author[11]{Timur Osipov}
\author[11]{Andrew Aquila}
\author[1,8,*]{N. Duane Loh}
\author[2,13]{Filipe R.N.C. Maia}
\author[3,5,14]{Pierre Thibault}
\affil[1]{Centre for BioImaging Sciences, National University of Singapore, Singapore 117557, {Singapore}}
\affil[2]{Laboratory of Molecular Biophysics, Department of Cell and Molecular Biology, Uppsala University, SE-75124 Uppsala, {Sweden}}
\affil[3]{Diamond Light Source, Harwell Science \& Innovation Campus, Didcot OX11 0DE, {United Kingdom}}
\affil[4]{Department of Physics \& Astronomy, University College London, London WC1E 6BT, {United Kingdom}}
\affil[5]{Department of Physics \& Astronomy, University of Southampton, Southampton SO17 1BJ, {UK}}
\affil[6]{Physical and Theoretical Chemistry Laboratory, Oxford University, Oxford OX1 3QZO, {United Kingdom}}
\affil[7]{European XFEL GmbH, Holzkoppel 4, 22869 Schenefeld, {Germany}}
\affil[8]{Department of Physics, National University of Singapore, Singapore 117557, {Singapore}}
\affil[9]{Division of Scientific Computing, Department of Information Technology, Science for Life Laboratory, Uppsala University, SE-75105 Uppsala, {Sweden}}
\affil[10]{Department of Applied Physics, Biomedical and X-Ray Physics, KTH – Royal Institute of Technology, SE 10691 Stockholm, Sweden}
\affil[11]{SLAC National Accelerator Laboratory, 2575 Sand Hill Road, Menlo Park,California 94025, {USA}}
\affil[12]{Brookhaven National Laboratory, Upton 11973 New York, {USA}}
\affil[13]{NERSC, Lawrence Berkeley National Laboratory, Berkeley, California 94720, USA}
\affil[14]{Universit\'a degli Studi di Trieste, Trieste 34127, {Italy}}

\affil[*]{Corresponding author: duaneloh@nus.edu.sg}

\begin{abstract}
A well-characterised wavefront is important for many X-ray free-electron laser
(XFEL) experiments, especially for single-particle imaging (SPI), where individual bio-molecules randomly sample a nanometer-region of highly-focused femtosecond pulses. 
We demonstrate high-resolution multiple-plane wavefront imaging of an ensemble of XFEL pulses, focused by Kirkpatrick-Baez (KB) mirrors, based on mixed-state ptychography, an approach letting us infer and reduce experimental sources of instability.
From the recovered wavefront profiles, we show that while local photon fluence correction is crucial and possible for SPI, a small diversity of phase-tilts likely has no impact.  
Our detailed characterisation will aid interpretation of data from past and future SPI experiments, 
and provides a basis for further improvements to experimental design and reconstruction algorithms. 
\end{abstract}
\setboolean{displaycopyright}{true}

\begin{document}
\maketitle

\section{Introduction}
The prediction that short, intense and coherent X-ray pulses could be used to determine structures based on single-shot diffraction data from isolated particles \cite{Neutze2000a} has been a major driving force in the realisation of X-ray free-electron lasers (XFELs). 
Over the past decade, a large number of such single-particle imaging (SPI) experiments were performed at XFELs, including structural studies on soot particles \cite{Loh2012}, silver particles \cite{Barke2015}, helium droplets \cite{Bernando2017},
cell organelles \cite{Hantke2014}, whole living cells \cite{VanderSchot2015} and virus particles via classical coherent diffractive imaging (CDI)  \cite{Seibert2011,Ekeberg2015,Munke2016,Daurer2017,Reddy2017,Hosseinizadeh2017,Rose2018,Lundholm2018,Bielecki2019,Sobolev2020}, in-flight holography \cite{Gorkhover2018} and fluctuation scattering \cite{Kurta2017,Pande2018b}.

These XFEL experiments depend on highly-focused and spatially-coherent X-ray pulses. 
Furthermore, knowing how the intensities and phases of these pulses develop through the optical focus at high resolution is helpful for SPI.
This knowledge helps us determine where the nanometer-sized particles should be injected to reproducibly give bright diffraction patterns.
The phase and intensity profiles differ between individual pulses because of the self-amplified stimulated emission (SASE) pulse-generation mechanism. 
While pulse-to-pulse intensity fluctuations are typically monitored by upstream gas detectors, the complex-valued beam profiles of the focused pulses are much less well-studied.
Without well-characterised beam profiles experimental design, instrument alignment, and data interpretation can be delayed or severely handicapped.

There are different approaches to X-ray wavefront characterisation at XFELs, including Shack-Hartmann sensors \cite{Keitel2016}, ablative imprints \cite{Chalupsky2011}, Young's double slit experiment \cite{Vartanyants2011}, grating interferometry \cite{Rutishauser2012,Schneider2018,Liu2018}, diffraction from aerosols \cite{Loh2013,Daurer2017} and coherent scattering speckle analysis \cite{Sikorski2015}. 
Ptychography, an imaging technique that is actively developed at synchrotron radiation facilities, has been used successfully for "at wavelength" metrology \cite{Kewish2010a,Vila-Comamala2011,Huang2013}.
Its potential has also been demonstrated for wavefront characterisation at XFELs \cite{Schropp2013,Pound2020,Sala2020}. 

Many of these wavefront characterisation experiments indicate that variations between pulses are significant, and even speculated to adversely impact SPI.
To study this impact definitively requires an experiment that directly measures both the spatial and temporal variations of the XFEL pulses at the foci used for SPI imaging. 
Here we describe a ptychographic wavefront-sensing experiment designed specifically to study this impact. To image these intense foci with ptychography, the pulses must be attenuated to keep the target undamaged, which leads to dimmer diffraction in a single pulse exposure.
To obtain a stable ptychographic reconstruction we must signal average many dim patterns together. 
However, the intrinsic variations between pulses or sub-micrometer mechanical vibrations anywhere along the beam path can reduce the speckle-contrast in these averaged patterns, thus jeopardizing the ptychographic reconstruction. 

These issues with beam diversity could be overcome with a mixed-state approach \cite{Thibault2013} to account for unavoidable experimental variations encoded in the diffraction data. Furthermore, we describe a single-pulse fitting procedure that infers hidden sources of variation, e.g. sample stage vibrations. 
A similar strategy could also be applied to related imaging methods such as electron ptychography, which often face similar important but hidden experimental variations (e.g. drift, aberrations) \cite{Chen2020}.

This mixed-state reconstruction strategy allowed us to characterise the full spatial intensity and phase profile of a focused XFEL beam and study their implications for SPI experiments, in particular the expected shot-by-shot variations in fluence and phase tilt. 
Through numerical propagation along the beam axis, we quantified the phases and intensities at 
different defocus positions and computed their effect on SPI data collection efficiency (i.e. hit-rate). 
Using this quantitative reconstruction of the complex-valued beam profile we simulated a large SPI data set under realistic experimental conditions.
A unique insight from this is the importance of a photon fluence correction in SPI reconstruction pipelines.

\section{Wavefront sensing at the LCLS}
\subsection{Data collection}
\begin{figure*}[htbp]
%\centering
\fbox{\includegraphics[width=2\columnwidth]{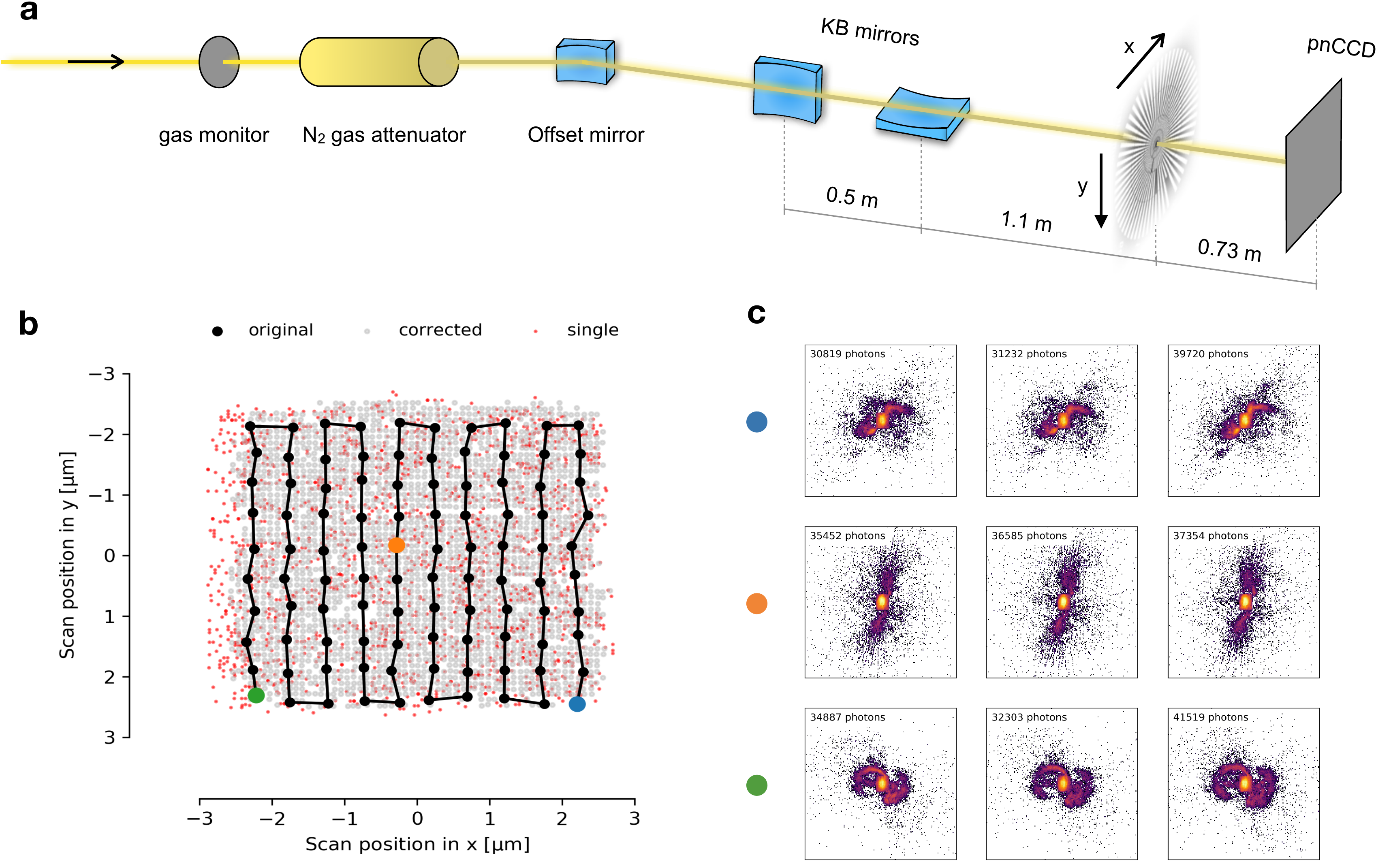}}
\caption{Experimental setup of the ptychographic wavefront sensing experiment (a). 
The attenuated AMO beam was focused onto a test pattern, scanning the beam in the transversal plane (b), and for each scan positions around $900$ diffraction patterns were recorded with a few examples shown in (c) on a log-scale. 
For the ptychographic reconstruction, the data has been prepared in $3$ different ways: using the nominal positions (black circles) averaging $300$ patterns, using combined corrected positions (grey dots) averaging at least $15$ patterns per position and using single corrected positions (red dots) without any averaging of the diffraction data. 
The ptychographic reconstructions for the $3$ types of data are shown in Fig. \ref{fig:recons}}
\label{fig:setup}
\end{figure*}

A ptychographic wavefront sensing experiment, with its setup shown in Fig. \ref{fig:setup}a, has been performed inside the LAMP endstation of the Atomic, Molecular and Optical Science (AMO) instrument \cite{Ferguson2015} at the Linac Coherent Light Source (LCLS) under beam conditions similar to SPI experiments.
A $\SI{200}{\nano\metre}$ thick Siemens Star test pattern, made by depositing gold (X30-30-2, \textcolor{blue}{www.zeiss.com}) on a $\SI{110}{\nano\metre}$ Si$_3$N$_4$ membrane, was used as a fixed target.
To limit damage to the target, the full and unfocused LCLS beam with a photon energy of $\SI{1.26}{\kilo\electronvolt}$ and an average pulse energy of $\SI[separate-uncertainty=true]{2.76\pm0.16}{\milli\joule}$ (as measured by an upstream gas monitor detector) was attenuated using a $\SI{4.26}{\metre}$ long nitrogen gas attenuator at an average pressure of $\SI{14.23}{}$ Torr resulting in a transmission of about $\SI{e-7}{}$. 
The attenuated beam was collimated using a pair of slits and focused onto the test pattern by a pair of Kirkpatrick-Baez (KB) mirrors which were located $\SI{1.1}{}$ and $\SI{1.6}{}$ m upstream of the interaction region.

The test pattern was scanned by moving the sample using a motorised piezo stage in a $10\times10$ "snake" pattern (black circles in Fig. \ref{fig:setup}b), while diffraction images were taken with a p–n junction charge-coupled device (pnCCD) detector \cite{Struder2010} 
placed $\SI{0.73}{\metre}$ downstream of the test pattern. 
The pnCCD was moved off-center to place the diffraction pattern on a $192\times192$ pixel area without any dead area. 
Due to using an upstream gas attenuator, no beamstop was necessary, and the signal on the detector was covering the complete dynamic range of diffraction from the sample. 
The scanning pattern was designed such that each position was given a randomised offset from a regular raster grid position 
to avoid raster grid pathology \cite{Thibault2009,Kewish2010a}. To reduce sample vibrations, the cryochillers of the pnCCDs were switched off temporarily during data acquisition. 
At each of the $\SI{100}{}$ scan positions $\mathbf{x}_j = (x_j,y_j)$, a total of $\SI{900}{}$ single-shot data frames were recorded distributed over $3$ separate data collections (each with $300$ frames per position). 
A selection of diffraction patterns is shown in Fig. \ref{fig:setup}c.

\subsection{Data pre-processing}
For each diffraction frame, a running dark subtraction was performed based on the average of the closest $100$ dark frames. 
Dark frames were recorded every two seconds using BYKIK, the upstream undulator beam kicker magnet. 
This dynamic thermal dark correction was necessary to account for changes in the readout of the pnCCD as it was slowly heating up. %This was a consequence of the pnCCD coolers being temporarily switched off to minimise vibrations on the sample stage.  
The detector was in gain mode $3$ where a photon at the given energy is equivalent to $25$ analogue-to-digital units (ADUs). 
For all detector  pixels, dark-corrected ADU values were converted to units of photons and values smaller than $0.8$ photons were rounded down to zero with all other values rounded to the closest integer value.

\section{Mixed-state reconstruction}
\subsection{Averaged data set with nominal positions}
For each of the $3$ data collections, all $300$ frames sharing the same nominal scanning position were averaged together to form a new set of ptychographic diffraction patterns $\hat{I}_{j\mathbf{q}}$, where $j$ is the position index and $\mathbf{q}$ is the reciprocal space coordinate. 
These averaged diffraction patterns have a much higher signal-to-noise ratio compared to individual frames but also a reduced speckle visibility due to pulse-to-pulse variations. This reduction step effectively produces a virtual data set, encoding any form of shot-to-shot variation in a "blurred" diffraction pattern.  
The mixed-state approach to ptychography \cite{Thibault2013} was designed precisely for such cases where one or more sources of partial coherence are present. 
In the mixed-state formalism, the diffracted intensity pattern at scan position $\mathbf{x}_j$ can be written as
\begin{equation}\label{mixed-state}
I_{j\mathbf{q}} = \sum_{m=0}^{M-1} \left| \mathcal{F}_{\mathbf{x}\rightarrow{}\mathbf{q}}\left[P^{(m)}_{\mathbf{x}-\mathbf{x}_j} O_\mathbf{x} \right]\right|^2
\end{equation}
where $M$ is a given number of probe modes (components) $P_\mathbf{x}^{(m)}$ in which the partially coherent illumination is decomposed, $O_\mathbf{x}$ the transmission function of the object and $\mathcal{F}_{\mathbf{x}\rightarrow{}\mathbf{q}} \left[\cdot\right]$ the 2D Fourier transform. 
The iterative reconstruction algorithm adapted for this problem, as described in \cite{Thibault2013} and implemented in \emph{PtyPy} \cite{Enders2016}, inverts \eqref{mixed-state} and simultaneously retrieves the object $O_\mathbf{x}$ 
and the $M$ probe modes $P_\mathbf{x}^{(m)}$ from the measured intensities $\hat{I}_{j\mathbf{q}}$ and positions $\mathbf{x}_j$.
At the end of each reconstruction, probe modes were orthogonalised and we calculated their relative power $w_m$.

\begin{figure*}[btp]
\centering
\fbox{\includegraphics[width=2\columnwidth]{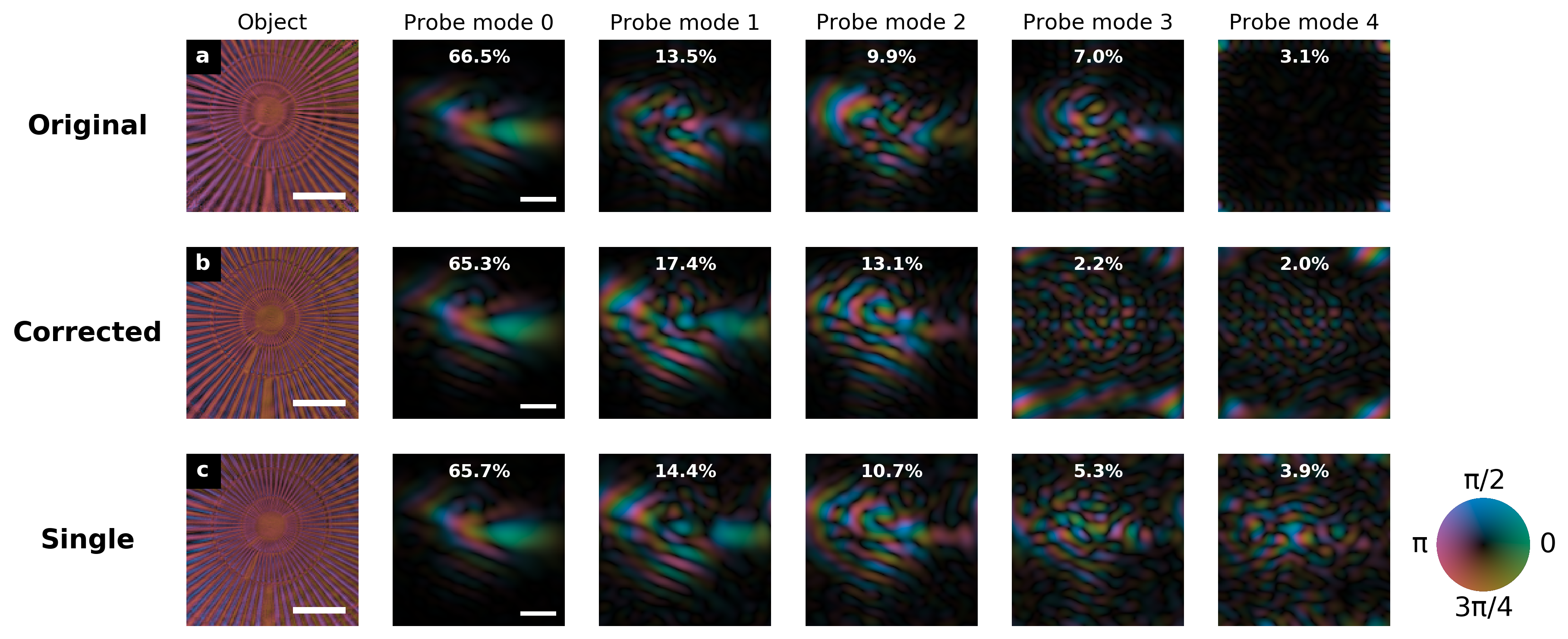}}
\caption{Three ptychographic reconstructions of a Siemens star test object together with $5$ incoherent probe modes.  
(a) One of the 3 original virtual data sets with $300$ frames averaged per scan position. 
(b) Corrected virtual data set with at least $15$ frames averaged per updated scan position and a total of $2642$ positions. 
(c) Single-frame data set with $2000$ updated scan positions. 
All images are complex-valued with phase mapped to color and amplitude mapped to hue. 
Scale bars are $\SI{2}{\micro \metre}$. 
Probe modes are orthogonal and scaled differently such that their full range of intensity is shown with the relative power $w_m$ of each mode written on top.}
\label{fig:recons}
\end{figure*}

We performed mixed-state ptychography reconstructions for each of the $3$ averaged virtual data sets and obtained the $M=\SI{5}{}$ most dominant probe modes, along with a high-resolution complex-valued image of the Siemens Star test pattern (Fig. \ref{fig:recons}a). 
We used $\SI{2000}{}$ iterations of the difference map (DM) algorithm \cite{Thibault2009} followed by $\SI{1000}{}$ iterations of the maximum-likelihood (ML) algorithm \cite{Thibault2012}. 
As an initial guess of the probe, we used an idealized illumination model based on the average $\SI{1.35}{\metre}$ focal distance between the two KB mirrors that gave a nominal focal size of $2 \times \SI{2}{\micro\metre\squared}$. 

Reconstructions with more than $5$ probe modes did not significantly change the relative power of the probe modes shown in Fig. \ref{fig:recons}a. Furthermore, the weakest reconstructed mode has distinct features forming towards the edges of the field of view. 
These features do not correspond to regions where the reconstruction has converged to the expected structure of the test patterns, and are attributed to possible detector readout artifacts in our measurement, an effect that had been previously observed \cite{Enders2014}.

\subsection{Checking for hidden sources of pulse-to-pulse variation}
Even though we have experimentally minimised vibrations on the sample stage, we still expected minor variations in the positions, either due to dynamic changes in the beam or the sample's location. 
XFELs are known to show some beam pointing instabilities that can lead to changes in the focused wavefront \cite{Emma2010}. 
In order to check for these types of pulse-to-pulse variations, we introduced a list of free parameters and define a model for a single-shot diffraction pattern $n$ as the incoherent sum
\begin{equation}
\begin{aligned}
I_{n\mathbf{q}} =  c_{n0} \left|\mathcal{F}_{\mathbf{x}\rightarrow{}\mathbf{q}}\left[P^{(0)}_{\mathbf{x}-\mathbf{x}_n-\Delta\mathbf{x}_n} O_{\mathbf{x}} A_{\mathbf{x}n} \right]\right|^2 + \\
\sum_{m=1}^{M-1} c_{nm} \left|\mathcal{F}_{\mathbf{x}\rightarrow{}\mathbf{q}}\left[P^{(m)}_{\mathbf{x}-\mathbf{x}_n} O_{\mathbf{x}} B_{\mathbf{x}n} \right]\right|^2 
\end{aligned}
\end{equation}
where $P^{(m)}_{\mathbf{x}}$ are the orthogonalised probe modes, $c_{nm}$ are positive real-valued coefficients, 
$\Delta\mathbf{x}_n = (\Delta x_n,\Delta y_n)$ is an additional translation of the main probe mode with respect 
to the object and $A_{\mathbf{x}n}$,  $B_{\mathbf{x}n}$ are additional phase terms for the
main and all other probe modes, respectively. These phase terms are defined as
\begin{align}
    A_{\mathbf{x}n} &= e^{i(\alpha_{nx}x + \alpha_{ny}y)} \quad \mathrm{and} \\
    B_{\mathbf{x}n} &= e^{i(\beta_{nx}x + \beta_{ny}y)}
\end{align}
with the phase differences $\alpha_{nx}$,  $\alpha_{ny}$,  $\beta_{nx}$ and  $\beta_{ny}$, 
which are related to phase tilt angles (see Section \ref{sec:tilt} in Methods) and further 
correspond to center shifts on the detector (and in the source plane). 
Altogether we have defined $M + 6$ scalars for a single pulse and 
treat them as fitting parameters in an optimisation scheme, 
minimizing the quadratic distance between the measured intensity pattern $\hat{I}_{n\mathbf{q}}$ and 
the model $I_{n\mathbf{q}}$. 
We used Powell's method, a conjugate direction approach without a need to calculate any derivatives, 
to obtain these single-shot parameters and cropped model and data to a $72\times72$ pixel area for improved 
computational efficiency. 

We performed this single-pulse fitting analysis on all $3$ data collections, with a total of $90$k single-shot events and discovered substantial variations in the positions with a root-mean-square-deviation (RMSD) 
from the nominal positions of 
$\SI{0.34}{\micro\metre}$ in the horizontal and $\SI{0.33}{\micro\metre}$ in the vertical direction.
Interestingly, the changing positions follow well-defined oscillations with a dominant frequency 
at around $\SI{30}{\hertz}$,
which can likely be attributed to electrical equipment in the vicinity of the sample chamber. We have also observed variations in the phase tilts and the coefficients, both are further discussed in the Results section. 

\subsection{Averaged data set with corrected positions}
With the corrected scan positions, we defined a new averaged virtual data set which combined the diffraction patterns from all $3$ data collections sampled on a much denser $60\times60$ grid with a total of $2642$ scan positions and at least $15$ patterns averaged per position (grey dots in Fig. \ref{fig:setup}b). Thereby, we have reduced the positional variation for each data point 
at the cost of lower signal-to-noise ratios. 
We performed another mixed-state reconstruction, again with a simulated initial guess for the probe, 
using $1000$ iterations of DM and $1000$ iterations of ML resulting in an improved reconstruction of the Siemens star together with the $5$ most dominant probe modes (Fig. \ref{fig:recons}b). 
The two weakest modes exhibit similar corner artefacts attributed to detector artefacts, as described above. 
We performed a second round of single-pulse fitting again on the full set of $90$k single-hit events but with the updated reconstruction and found that variations in the positions were reduced to a RMSD of 
$\SI{0.24}{\micro\metre}$ in the horizontal and $\SI{0.14}{\micro\metre}$ in the vertical direction compared to the first round. Further rounds of single-pulse fitting did not significantly improve the corrected positions.  

\subsection{Single-frame data set with corrected positions}
For the final ptychographic reconstruction, we selected $20$ single-shot events per 
nominal position, giving a total of $2000$ events each with corrected positions
(red dots in Fig. \ref{fig:setup}b). We then performed a mixed-state reconstruction using $2000$ iterations of the ML algorithm 
with an iteration-dependent smoothing operation on the object. Starting with $20$ pixels, the size 
of the Gaussian smoothing kernel was reduced exponentially with a decay rate of $1/200$ per iteration. As the initial guess
for the illumination, we used the $5$ probe modes from the previous reconstruction (Fig. \ref{fig:recons}b) and started 
updating the probe modes after $500$ iterations. The final reconstruction of the Siemens star, along with the $5$ most dominant 
modes is presented in Fig. \ref{fig:recons}c. Based on the given geometry of the experiment, the half-period resolution (pixel size) was $\SI{50.3}{\nano\metre}$. 

\section{Results}
\begin{figure}[tb]
\centering
\fbox{\includegraphics[width=0.95\columnwidth]{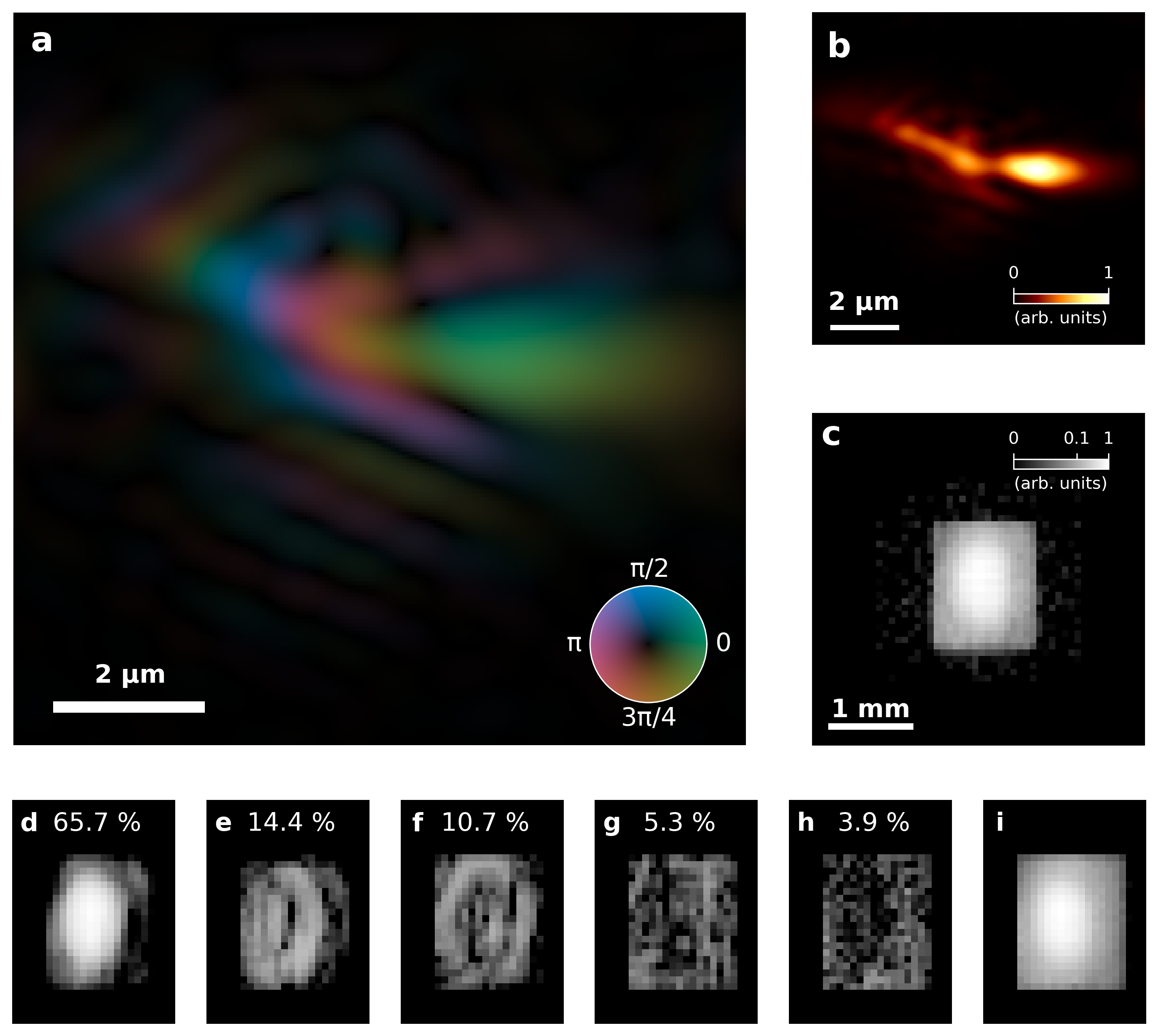}}
\caption{Ptychographic wavefront reconstruction of the average AMO beam. 
(a) The most dominant probe mode with a power of $65.7\%$. Phase is mapped to color, amplitude mapped to hue. 
(b) The incoherent sum of all $5$ probe modes estimating the average fluence profile at AMO. 
(c) The recorded direct beam intensity of a single pulse on a log scale in the same configuration as the ptychographic experiment.
(d-h) Intensity of the back-propagated probe modes from Fig. \ref{fig:recons}c on a log scale with their relative powers $w_m$ given as percentages. 
(i) Incoherent sum of the back-propagated probe modes on a log scale. Images (d-i) share the same gray scale as (c).}
\label{fig:beam}
\end{figure}

After two rounds of data averaging and single-pulse fitting, we identified the $5$ reconstructed modes (Fig. \ref{fig:recons}c) 
as the best approximation of the partially coherent average AMO beam, with $\SI{65.7}{\percent}$ of the total power in its most dominant mode (Fig. \ref{fig:beam}a). We note that partial coherence 
in this context describes all sources of "effective" decoherence, including unresolved fringes and speckles produced by scatterers along the optical path, and detector point-spread-function. 
However, back-propagating each orthogonal probe mode into the secondary source plane and inspecting its 
intensity (Fig. \ref{fig:beam}d-h) suggests that all $5$ modes, incoherently summed together 
(Fig. \ref{fig:beam}i), are necessary to describe the features in the recorded direct beam intensity of a 
single LCLS pulse focused by the KB optics at AMO (Fig. \ref{fig:beam}c). 
Therefore, we present the average intensity profile at AMO as the incoherent sum of the $5$ probe modes (Fig. \ref{fig:beam})
measured by mixed-stated ptychography $\SI{0.5}{\milli\metre}$ upstream of the focal plane with the strongest intensity 
(as determined by numerical propagation). 

The most intense region of the focus has an estimated full width at half maximum (FWHM) of $\SI{2}{\micro\metre}$ in the horizontal and $\SI{0.5}{\micro\metre}$ in the vertical direction with a large tail extending several micrometres towards the upper left corner. 
In the same intense focus region, the phase in the average wavefront appears to be flat with small changes towards the edges as the intensity gets weaker and large changes in the tails of the beam. 
These tails (or "wings") in a KB-focused beam are often attributed to mid-spatial frequency roughness of the mirror \cite{Soufli2008,Pivovaroff2007}.  

In the following sections, we derive multiple properties of the reconstructed intensity and phase profiles numerically propagated into different defocus planes (see Section \ref{sec:propagation} in Methods for details) and describe their implications on single-particle imaging. 

\subsection{The beam intensity profile seen by single particles}
\begin{figure*}[btp]
\centering
\fbox{\includegraphics[width=2\columnwidth]{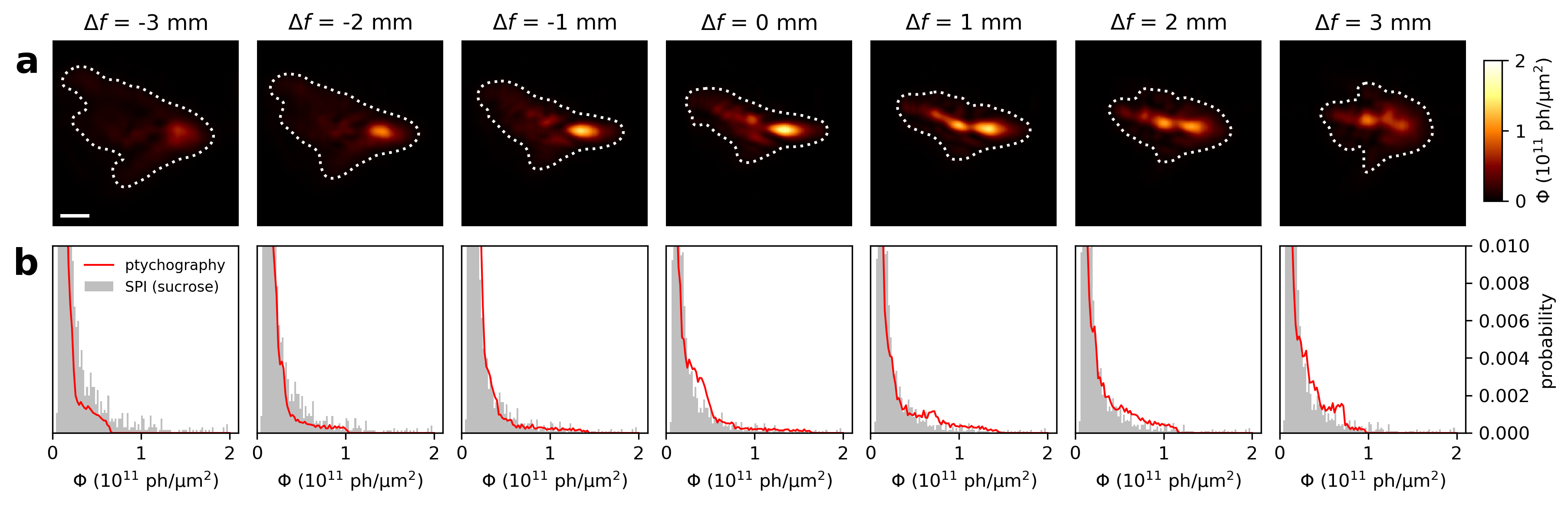}}
\caption{Fluence profiles at different defocus $\pm \SI{3}{\milli\metre}$ from the plane with the 
strongest observed peak fluence. 
(a) Estimated full beam fluence at $\SI{2.76}{\milli\joule}$ average pulse energy and $4.8\%$ transmission. 
The white-dotted contour indicates $\SI[per-mode=symbol]{5e9}{\photons\per\micro\metre\squared}$.
(b) Fluence distributions, comparing the histogram of (a) to the distribution obtained from single-particle 
sucrose diffraction data. The distributions obtained by ptychography (red curve) are normalised such that their integral 
equals to 1. The distribution obtained by sucrose SPI is rescaled to match the average pulse energy of $\SI{2.76}{\milli\joule}$ 
and normalised such that the integral is equal to the ptychography distribution for fluences above 
$\SI[per-mode=symbol]{5e9}{\photons\per\micro\metre\squared}$. 
The bin width is $\SI[per-mode=symbol]{2e9}{\photons\per\micro\metre\squared}$. }
\label{fig:defocus}
\end{figure*}

Our recovered attenuated photon fluence distribution, when properly rescaled (Fig. \ref{fig:defocus}), is comparable to an independent study performed at the same beamline with sucrose particles \cite{Ho2020}.
For upscaling the recovered intensity profile at different defocus planes, 
we used an unattenuated pulse energy of $\SI{2.76}{\milli\joule}$ (equivalent to $\SI{1.38e13}{}$ photons) and an estimated transmission through the optics of $\SI[separate-uncertainty=true]{4.8\pm0.4}{\percent}$ 
(see Section 1 in \textcolor{blue}{Supplement 1} for details). 
This transmission efficiency is reasonable when compared to more recent power measurements \cite{Heimann2019} given the additional apertures used in our experiment (see Fig. \ref{fig:beam}i). 
The upscaling approximates the fluence distributions as observed by a single particle in a regular SPI experiment at AMO. 
While 2D profiles as shown in Fig. \ref{fig:defocus}a are typically unknown in such experiments, it is possible to 
map out their 1D distributions (Fig. \ref{fig:defocus}b) based on a fitting analysis of single-shot diffraction from 
particles randomly sampling the beam \cite{Loh2013,Hantke2014,Daurer2017,Ho2020,Sobolev2020}.  
Fig. \ref{fig:defocus}b shows that the distributions obtained from this ptychographic experiment (red curves) agrees with the fluence histograms (shown in gray) from sucrose cluster diffraction obtained at a slightly lower photon energy ($\SI{1140}{\electronvolt}$) and pulse energy ($\SI{1.1}{\milli\joule}$) but otherwise similar conditions at the AMO endstation \cite{Ho2020,CXIDB119}. The sucrose data has been rescaled to match the average pulse energy of the ptychography data and adjusted to consider the $\SI{70}{\percent}$ dynamic scattering efficiency reported in \cite{Ho2020}.  
The agreement between the two measurements persists within $\pm 3$ mm of the nominal focus. 

\subsection{The "optimal" defocus position for single-particle imaging}
One of the most time-consuming challenges in SPI is to adjust the overlap between the particle beam and the X-ray pulses to maximise the fluence and the ratio of detected single-particle events (hit-ratio) \cite{Aquila2015,Hantke2014}. 
With the reconstructed 2D fluence profile, we can also define this hit-ratio as the area of all spots with a fluence larger than a certain detection limit multiplied by the particle density 
(particles per area). 
This particle density is dependent on particle delivery conditions, here we assume $\SI[per-mode=symbol]{0.001}{\particles\per\micro\metre\squared}$ which seems like a reasonable choice for commonly used particle injectors \cite{Hantke2018}.

\begin{figure}[tbp]
\centering
\fbox{\includegraphics[width=0.95\columnwidth]{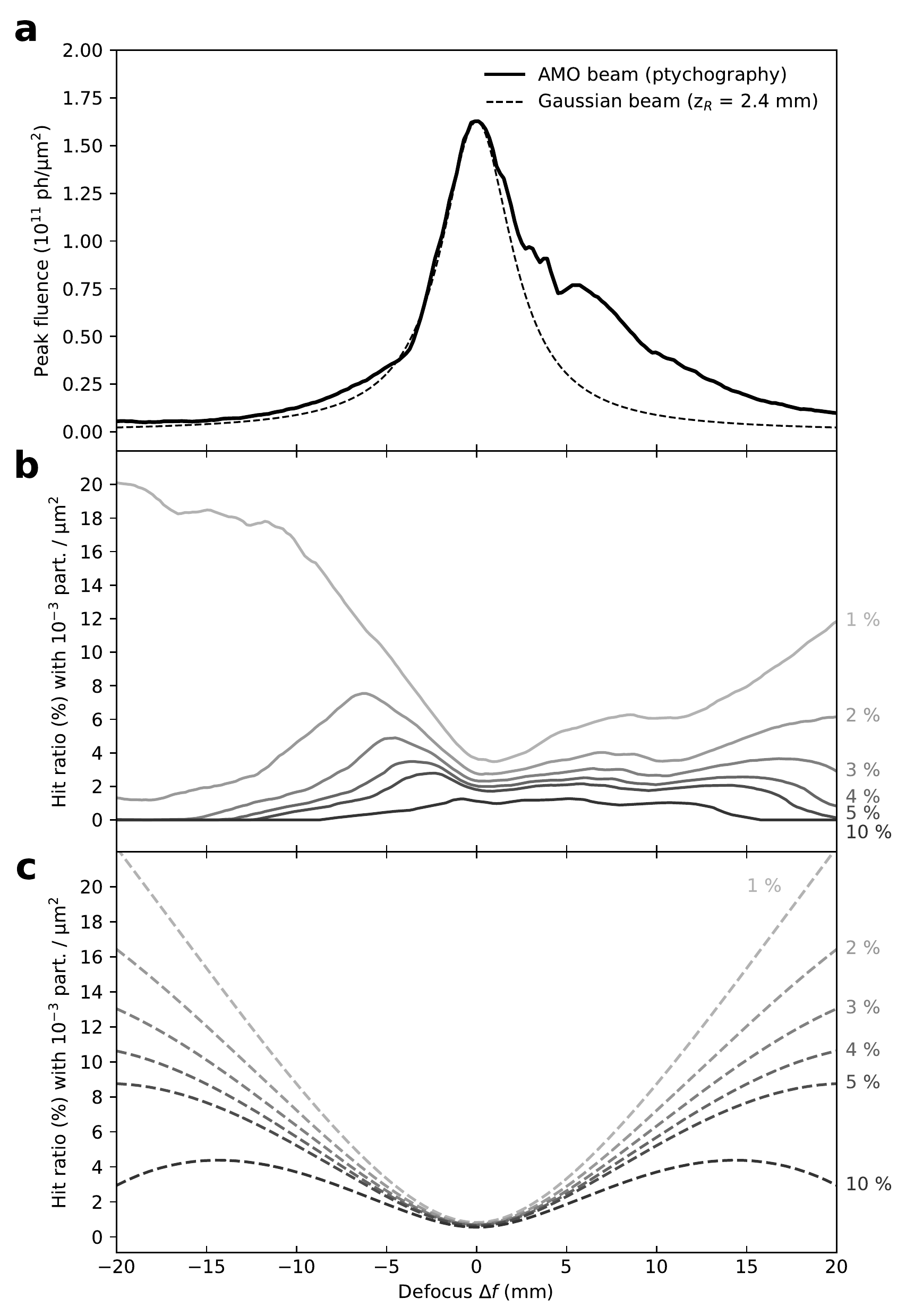}}
\caption{Beam properties in a wide range of defocus $\Delta f$ from $\SI{-20}{\milli\metre}$ to $\SI{20}{\milli\metre}$. 
(a) Peak fluence as obtained from the propagated AMO fluence distributions (solid) compared to the peak fluence expected 
from a Gaussian beam matching the peak fluence at $\Delta f=\SI{0}{\milli\metre}$ and with a Rayleigh length of 
$\SI{2.4}{\milli\metre}$.
(b) Expected single-particle hit ratios derived from the propagated fluence distributions  
assuming a particle density of $\SI[per-mode=symbol]{0.001}{\particles\per\micro\metre\squared}$ 
and shown for detection limits at different fractions of the peak fluence at $\Delta f=\SI{0}{\milli\metre}$. 
(c) Expected single-particle hit ratios for a Gaussian beam under the same conditions as described in (b).}
\label{fig:hitrate}
\end{figure}

In practice, one has to find the optimal possible trade-off between these two quantities (fluence and hit-ratio) as the focal plane with the highest peak fluence (the focus) usually also minimises the achievable hit ratio (Fig. \ref{fig:hitrate}). 
Similarly, going further away from the focus increases the hit ratio while decreasing the peak fluence. 
This is always the case in Gaussian beam optics no matter which level of fluence is used for defining the single-particle hit detection limit (Figs. \ref{fig:hitrate}a and \ref{fig:hitrate}c).

Looking at how the peak fluence changes with defocus (Fig. \ref{fig:hitrate}a), we can see that the KB-focused beam at AMO is similar to a Gaussian beam with Rayleigh length $\SI{2.4}{\milli\metre}$ but with highly asymmetric features downstream of the focus. 
Similar effects have been found in ablation studies of the same KB-focused beam \cite{Chalupsky2011}. 
Furthermore, we also see a similar trend in the expected hit-ratios compared to a Gaussian beam, but crucially as the hit detection limit increases, which is equivalent to imaging smaller and smaller particles, the situation is different and the position along the beam axis
with the minimal hit ratio is no longer found at the focus (Figs. \ref{fig:hitrate}a and \ref{fig:hitrate}c).
This also means that measuring the hit ratio as the particle stream is scanned
along the X-ray beam axis is likely to give a biased estimate of the focal plane.

Concerning the 
best choice of defocal plane for SPI in the given KB-focused beam at AMO (or other similar KB-focused beams), there is 
no obvious answer because it highly depends on the size and detectability of the particles. At a given particle density 
of $\SI[per-mode=symbol]{0.001}{\particles\per\micro\metre\squared}$, larger particles that can be detected 
as low as only 
$\SI{1}{\percent}$ of the maximum achievable peak fluence might still give rise to reasonable diffraction intensities 
at $\SI{-10}{\milli\metre}$ defocus with much smaller peak fluence but also a much higher hit ratio. Smaller particles 
that are only detectable above $\SI{10}{\percent}$ of the maximum achievable peak fluence should probably be 
imaged closer to the focus where both fluence and hit ratio can be  maximised 
(Figs. \ref{fig:hitrate}a and \ref{fig:hitrate}b). 
 
\subsection{Strong fluence variations impact structural inference}

\begin{figure*}[btp]
\centering
\fbox{\includegraphics[width=1.95\columnwidth]{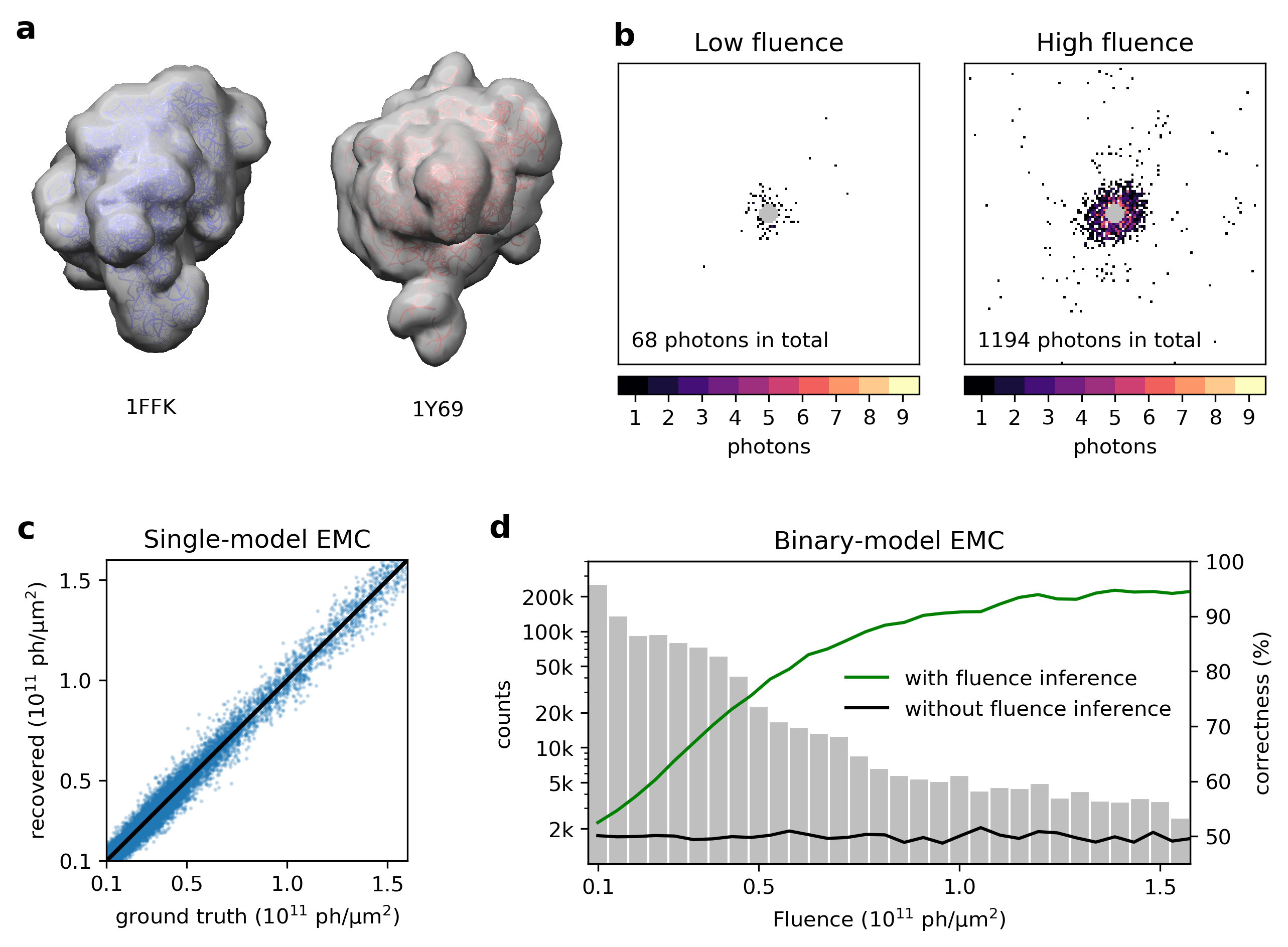}}
\caption{Simulated single-particle reconstructions using EMC. 
(a) 1FFK and 1Y69, two similar but different $\SI{1.4}{MDa}$ structures from the Protein Data Base (PDB) with their electron densities at $\SI{1.5}{\nano\metre}$ resolution rendered in gray using Chimera \cite{Pettersen2004}. 
(b) Example diffraction from 1FFK in a random orientation at low ($\SI[per-mode=symbol]{1e10}{\photons\per\micro\metre\squared}$) 
and high ($\SI[per-mode=symbol]{16e10}{\photons\per\micro\metre\squared}$) fluence.
(c) Comparison of the recovered and true fluence when running regular (single-model) EMC with $1$ million patterns generated 
from structure 1FFK at relevant AMO parameters and using the expected in-focus 
fluence distribution (Fig. \ref{fig:defocus}a). The black line indicates equal values for true and recovered fluence.
(d) Two-model (binary) EMC with $500$k 1FFK patterns and $500$k 1Y69 patterns, highlighting 
the importance of fluence correction in multi-model single-particle reconstructions. 
The correctness score is defined as the fraction of patterns correctly classified into the two models for 1FFK and 1Y69.
The underlying fluence distribution with a bin size of $\SI[per-mode=symbol]{5e9}{\photons\per\micro\metre\squared}$ 
is provided for reference in gray on a log scale.}
\label{fig:emc}
\end{figure*}

To reconstruct the 3D Fourier intensity in SPI, is equivalent to inferring each particle's unknown orientation \cite{Loh2009} and the local incident photon fluence \cite{Loh2010,Ekeberg2015} from only their respective diffraction measurements. 
Our reconstructed pulse profile in Fig. \ref{fig:defocus} clearly demonstrates how this local fluence can vary by orders of magnitude depending on where injected particles randomly intercept the focused pulse, rather than pulse-to-pulse variations from the SASE lasing process.
While such fluence variation between patterns is sometimes corrected as a pre-processing step in SPI by normalizing the number of photons per pattern, this is challenging when the patterns are severely photon-limited Fig. \ref{fig:emc}b.

Here, we used the reconstructed average pulse profile to study if the variations in local photon fluences for different illuminated particles can be correctly inferred, and the impact of such variations on structure determination. 
With the \emph{Dragonfly} package \cite{Ayyer2016} we simulated $1$ million single-particle diffraction patterns of $\SI{1.4}{MDa}$ particles (Fig. \ref{fig:emc}a,b) randomly placed in the expected beam profile at $\Delta f = \SI{0}{\milli\metre}$ (Fig. \ref{fig:defocus}). 
For this simulation, we used relevant AMO parameters with a photon energy of $\SI{1.26}{\kilo\electronvolt}$ and the pnCCD detector placed at $\SI{130}{\milli\metre}$ behind the interaction region with a pixel pitch of $\SI{75}{\micro\metre}$ and a circular beamstop of radius $\SI{30}{px}$. 
A `hit' detection limit of $\SI[per-mode=symbol]{1e10}{\photons\per\micro\metre\squared}$, 
corresponding to around $50$ scattered photons per pattern, was assumed.

For a single structural class, we found that the expand-maximize-compress (EMC) \cite{Loh2009} SPI reconstruction algorithm (see Section \ref{sec:emc} in Methods for details) adequately recovers the ground truth photon fluences of each particle/pattern (Fig. \ref{fig:emc}c). 

Surprisingly, if the patterns were instead randomly generated from two different $\SI{1.4}{MDa}$ structures (Fig. \ref{fig:emc}a), Fig. \ref{fig:emc}d shows that the correct structural class cannot be determined without also inferring (hence correcting for) the fluence variations. 
Furthermore, the structural classes are only correctly inferred for particles illuminated by sufficiently high local photon fluences ($\gtrapprox 5\times 10^{10}$ ph/$\mu$m$^2$).
This dependence on fluence supports the importance of the few photons scattered to sufficiently high angles, which are important for distinguishing structural classes through their higher-resolution difference.

%After $70$ iterations of regular single-model EMC with \emph{Dragonfly} (see Section \ref{sec:emc} in Methods for details), 

%To further prove the importance of fluence correction, we have done another simulation mixing diffraction data from two slightly different $\SI{1.4}{MDa}$ structures (Fig. \ref{fig:emc}a) and used binary-model EMC (see Section \ref{sec:emc} in Methods for details) to not only recover the orientation of the particles but also classify them into two expected 3D Fourier intensity models. We again simulated $1$M diffraction patterns, $500$k for each structure under the same experimental conditions as described above. 
%After $180$ iterations of binary-model EMC with \emph{Dragonfly} we found that without using fluence correction, the binary-model classification is only as good as randomly assigning diffraction patterns to either of the two models ($\SI{50}{\percent}$ correctness). 
%With fluence correction turned on, the fraction of correctly assigned patterns was always bigger than $\SI{50}{\percent}$ reaching $\SI{95}{\percent}$ for the patterns with the highest incident fluence (Fig. \ref{fig:emc}d).

\subsection{The impact of phase tilts on single-particle diffraction}
Besides the optimisation for fluence and hit ratios, in SPI it is also desired to inject the particles into a flat wavefront \cite{Aquila2015}. 
Several single-particle studies have observed displacements in the centre of diffraction attributed to differences in phase tilts between local regions of different pulse wavefronts intersected by different particles \cite{Loh2013,Daurer2017,Sobolev2020}. 
These variations in the phase tilt can either be due to shot-to-shot fluctuations of the wavefront (e.g. beam pointing) or because of local 
variations in the phase profile of the beam sampled by the randomly positioned particles. 
These two sources of variations, unfortunately, are indistinguishable in an SPI experiment. 
Our ptychographic pulse profiling allows us to directly measure these two sources of phase tilt variations, and study their independent impact on SPI.

When fitting our ptychographic model to single-pulse diffraction data, we found that the changes in the overall phase tilt between pulses for the dominant mode have a small standard deviation of $\SI{87}{\micro\radian}$ and $\SI{32}{\micro\radian}$ for the horizontal and vertical components respectively. 
For all other modes combined, these standard deviations shrunk to $\SI{8}{\micro\radian}$ and $\SI{6}{\micro\radian}$ respectively.
For a detector distance of $\SI{130}{\milli\metre}$ (typical for SPI experiments at AMO) this would translate into displacements of the centre of diffraction up to $\SI{11}{\micro\metre}$, which are much smaller than the pixel pitch of the pnCCD detector. 
This suggests that the overall pulse-to-pulse wavefront fluctuations in the LCLS wavefront should only cause negligible displacements in the centre of diffraction often observed in SPI.

\begin{figure*}[htbp]
\fbox{\includegraphics[width=2\columnwidth]{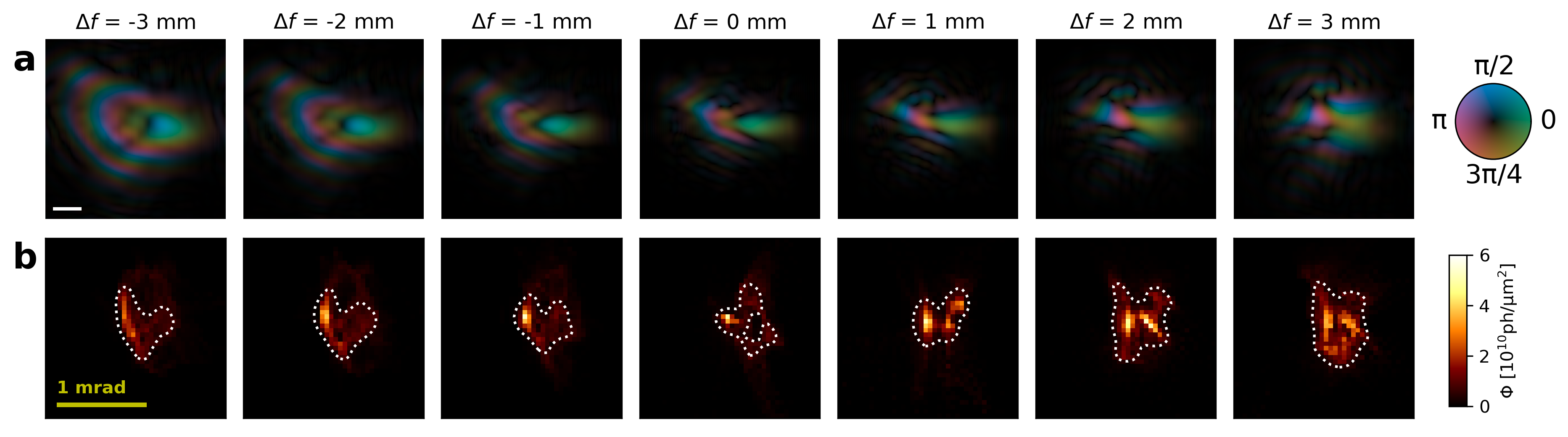}}
\caption{Phase profiles at different defocus $\pm \SI{3}{\milli\metre}$ from the plane with the 
strongest observed peak fluence. 
(a) Most dominant probe mode, phases are mapped to color, amplitudes are mapped to hue. The scale bar is $\SI{2}{\micro\metre}$
(b) A mapping of the local phase tilts based on the profile in (a) with $40\times40$ bins of size $\SI{0.05}{\milli\radian}$. 
The color refers to the average fluence in each bin ($\SI{2.7}{\milli\joule}$ average pulse energy, $4.8\%$ transmission). 
The white-dotted contour indicates $\SI[per-mode=symbol]{5e9}{\photons\per\micro\metre\squared}$.}
\label{fig:phase}
\end{figure*}

Since the differences in overall phase tilts between pulses are small, the random shifts observed in SPI diffraction patterns of different particles might be due to where each particle intersects a curved (and possibly irregular) pulse wavefront. 
To study this second source of local phase tilt variation, we computed the gradient in the horizontal and vertical direction of the most dominant pulse mode in our ptychographic reconstruction (shown at different defocus planes in Fig. \ref{fig:phase}a). 
We again converted those pixel-to-pixel phase differences to a distribution of phase tilts (see Section \ref{sec:tilt} in Methods for details) that a single particle would observe in an SPI experiment (Fig. \ref{fig:phase}b) with the average fluence mapped in color. 
The white-dotted contour lines indicate an average fluence of $\SI[per-mode=symbol]{5e9}{\photons\per\micro\metre\squared}$. 

The fluence-phase tilt distribution that we reconstructed here at $\SI{3}{\milli\metre}$ defocus resembles that reported earlier (Fig. 4a of ref. \cite{Loh2013}). 
Considering all defocus planes shown in Fig. \ref{fig:phase}, the expected deviation is less than $\SI{1}{\milli\radian}$ translating to a $\SI{130}{\micro\metre}$ displacement  at $\SI{130}{\milli\metre}$ detector distance. 

Overall, local variations in phase tilts on a single pulse's wavefront are larger than variations in overall phase tilts between pulses. 
Nevertheless, for sufficiently small particles whose diffraction speckles are well sampled at low scattering angles, this larger source of phase tilt variations is unlikely to have a big impact on SPI experiments at AMO.
Furthermore, these small displacements in the centre of diffraction can be inferred and corrected when individual patterns contain sufficiently many photons \cite{Daurer2017}.

\subsection{Correlation with electron bunch and pulse properties}
Each of the $90$k pulses used in this ptychographic experiment includes a set of diagnostic EBEAM parameters that measure properties of each electron bunch and photon pulse along their trajectory in the accelerator tunnel and inside the optics hutches respectively. 
A comprehensive description of these parameters can be found in \cite{Sanchez-Gonzalez2017}.

We correlated these EBEAM parameters with the parameters fitted for individual pulses in our ptychographic reconstruction pipeline. 
For this purpose, we have chosen the robust non-parametric Spearman's rank correlation.
Most notably, we found that only the equatorial angular "wobble" of individual pulses ($\alpha_x$) that was inferred from our ptychographic reconstructions seems to follow fluctuations in a subset of upstream electron bunch properties. 
Although other reconstructed parameters, such as $c_0$ and $c_2$, also showed significant correlations with EBEAM parameters (see Section 3 and Table S1 in \textcolor{blue}{Supplement 1}), their physical significance is less obvious. 

Importantly, inferred simple translations of individual pulses (e.g. $\Delta x$ and $\Delta y$) show little to no correlation with EBEAM parameters. 
Hence, it seems plausible that most of these translational variations might be attributed to vibrations of the sample stage, the KB optics or the offset mirror rather than beam pointing instabilities further upstream. Coincidentally, the translational variations in our reconstructions (see supplementary Fig. S2) have characteristic frequencies that are most likely caused by equipment close to the sample stage.

We anticipate that future beam optimisation experiments at XFELs in combination with machine-learning approaches \cite{Sanchez-Gonzalez2017} where the focused beam under relevant SPI conditions is monitored (e.g. using single-pulse ptychography \cite{Sala2020}) while adjusting the upstream beam parameters would allow for smart pulse-picking strategies that could potentially improve the data quality for SPI. 

\section{Discussion}
%We have characterised the KB-focused beam at the AMO endstation at the LCLS under conditions relevant for SPI and other single-shot XFEL experiments. 
In addition to their high brilliance, XFEL pulses are prized for SPI because of their high spatiotemporal coherence.
Naturally, several prior experiments were dedicated to measuring this degree of coherence \cite{Vartanyants2011,Lee2020}.
While these measurements relied on the assumption that the pulses comprised a few low order Hermite-Gaussian modes, our mixed-state ptychographic wavefront sensing directly decomposes the ensemble of XFEL pulses into a set of orthogonal wavefront modes without such assumptions (Fig. \ref{fig:beam}).
The power distribution of our recovered modes gives a useful quantification of the coherence properties of the beam.
We have found that on average $\SI{65.7}{\percent}$ of the total power was present in the dominant mode of the beam, which means that an hypothetical ideal spatial filter could create a fully coherent beam with about $65\%$ of the pulse fluence. This relative power of the dominant mode is lower than the $\SI{78}{\percent}$ previously reported for the neighbouring SXR endstation at the LCLS \cite{Vartanyants2011}, though the widely differing experimental settings prevent a direct comparison of these numbers.
Another metric is the degree of coherence \cite{Born1999}, which can simply be computed by summing the squares of the mode relative powers $w_m$:
\begin{equation}\label{degree-of-coherence}
%\xi = \sum_{m=0}^{M-1} \left( \int \left| P^{(m)}_{\mathbf{x}} \right|^2 d\mathbf{x} \right)^2
\xi = \sum_{m=0}^{M-1} w_m^2.
\end{equation}
We find a degree of coherence of $\xi= \SI{46.8}{\percent}$, again lower than the value of $\SI{56}{\percent}$ previously reported for the SXR endstation. While our method does not discriminate between the possible sources of coherence loss, it is likely that scattering from upstream optics and detector point-spread-function are the main factors.

Unlike fixed target experiments, the random injection of small particles into the stream of XFEL pulses in SPI has decidedly an element of chance and uncertainty.
Our mode decomposition of these pulses provides a unique window to efficiently explore how random samples of different regions of many pulses can impact on SPI.

To first order, we found that previously reported random translations in the diffraction centers of individual diffraction patterns \cite{Loh2013,Daurer2017} can be largely explained by particles randomly sampling different regions of pulses with irregular local phase tilts (Fig. \ref{fig:phase}).
Such irregularity is primarily determined by the surface finish of the KB-focusing mirrors \cite{Barty2009}.
Our pulse-to-pulse regressions showed that the variations in overall phase tilt between pulses played a lesser role here.
Within a few millimeters of the pulses' foci, we found the maximum observed phase tilt was only around $\SI{1}{\milli\radian}$, which in most scenarios only has a minimal influence on SPI data analysis.

In contrast, the variations in local photon fluence as particles randomly sample different spots in the focused LCLS beam has a much larger impact on SPI.
These variations were previously observed indirectly from randomly injected particles at both the AMO \cite{Loh2013,Hantke2014,Ho2020} and CXI beamlines \cite{Daurer2017} at the LCLS, and recently also at the SPB instrument at the European XFEL \cite{Sobolev2020}. 
Ptychographic reconstructions, however, allowed us to directly observe how these spatial intensities distributions develop near the pulses' nominal foci (Fig. \ref{fig:defocus}).
Depending on where the particle beam intersects the X-ray beam along its propagation axis, the intensity profile also has interesting and non-intuitive implications on the expected hit ratio (Fig. \ref{fig:hitrate}).

Comprehensive beam profiling experiments help us better anticipate, measure, and plan and/or correct for uncertainties in an SPI experiment that may degrade imaging resolution. 
In this capacity, ptychographic wavefront sensing using a mixed-state approach will be useful for complex SPI experiments where perfect knowledge of all parameters is practically impossible (e.g. sources of vibrations in Fig. \ref{fig:recons} ). 
The importance of robust photon fluence correction for 3D SPI reconstruction algorithms and structural inference that is demonstrated in Fig. \ref{fig:emc}, is one such example.

Rapidly profiling the foci of tens of thousands of pulses during an SPI beamtime using ptychography can also help save precious experiment time. 
In principle, collecting and reconstructing the wavefront modes of 10,000 pulses at 120 Hz repetition rate may only take minutes, which gives timely feedback for tuning optical elements to maximize photon transmission, changing focus conditions, and/or reducing background scattering. 
Currently, such tuning because of limited fast feedback can take several hours, if such tuning is at all successful.
A routine and well-optimised ptychographic pulse profiling instrument can substantially reduce such optical tuning times.
Furthermore, such profiles can rapidly form estimates of ideal and target hit rates (Fig. \ref{fig:beam}), which in turn allow experimenters to efficiently diagnose potential issues with particle injectors (e.g. clogging).

Overall, we are confident that live pulse profiling will be important for efficient SPI experiments.
This article demonstrates that mixed-state ptychography can serve this key role.
Other XFEL experiments that assemble insights from many partial measurements of an ensemble of pulses may also benefit from such detailed profiling.

\section{Methods}

\subsection{Numerical wavefront and intensity propagation}\label{sec:propagation}
Based on the orthogonalised probe modes $P^{(m)}_{\mathbf{x}}$ and the wavelength $\lambda$, we can define the propagated
probe at a near distance $z$ as 
\begin{equation}
    P^{(m)}_{\mathbf{x}}(z) = \mathcal{F}^{-1}_{\mathbf{q}\rightarrow{}\mathbf{x}} \left[
     \mathcal{F}_{\mathbf{x}\rightarrow{}\mathbf{q}} \left[ P^{(m)}_{\mathbf{x}} \right] 
     \exp \left[ \frac{i 2\pi z}{\lambda} \left(\sqrt{1 - \mathbf{q}^2\lambda^2} - 1 \right)\right] 
     \right] \,,
\end{equation}
using the numerically stable and convenient angular spectrum method, neglecting the plane wave term.
This allows us to further define fluence maps $\Phi_\mathbf{x}(z) = \sum_m |P^{(m)}_{\mathbf{x}}(z)|^2$ 
as the incoherent sum of the $M$ propagated probe modes. 

\subsection{The binary-model EMC algorithm}\label{sec:emc}
The single-model EMC reconstruction in Fig. \ref{fig:emc}c iteratively maximises the target log-likelihood function in 
\cite{Loh2009, Loh2010, Ayyer2016},
\begin{equation}
    Q(W',\phi', W, \varphi) = \sum_{d, r, t} P_{dr}(K_d|W_r \varphi_d)\bigl[K_{dt} \log(W'_{rt}\varphi_d') - W'_{rt}\varphi_d'\big]\; ,
    \label{eq:dragonfly_Q}
\end{equation}
where the current model for the diffraction intensities $W$ is updated to $W'$, and the likely local photon fluence ($\varphi_d$) that illuminated the particle that resulted in diffraction pattern $K_d$ updated to $\varphi_d'$.
Here, the set of diffraction patterns are denoted $K_{dt}$, where the $d$ subscript denotes the pattern index, and $t$ subscript defines the pixel index on each pattern. 
The maximisation of \eqref{eq:dragonfly_Q} runs over a set of discrete samples of the 3D rotation group, indexed $r$, based on a linear refinement scheme of the unit quaternions that correspond to the 4D 600-cell \cite{Loh2009}. 
The likelihood function in \eqref{eq:dragonfly_Q} is
\begin{equation}
    P_{dr} (K_d | W_r \varphi_d) = \prod_t \exp{(- W_{rt} \varphi_d)} (W_{rt}\varphi_d)^{K_{dt}}/K_{dt}! \; ,  
\end{equation}
where $W_{rt}$ is the Ewald sphere section of the diffraction volume $W$ at orientation $r$ that is sampled by detector pixels labeled $t$. 
We maximise the log-likelihood in \eqref{eq:dragonfly_Q} by solving $\text{d} Q/ \text{d} W_{x}=0$ and $\text{d}Q/\text{d}\varphi_d=0$ alternately, where $x$ is the voxel index of the Fourier intensity volume. 
This gives the typical update of the intensity volume and local photon fluence:
\begin{align}
    W'_{x} &= \dfrac{\sum_{d} \sum_{\{rt;x\}}P_{dr}K_{dt}}{\sum_d
    \sum_{\{rt;x\}}P_{dr}\varphi_d}\text{,} \\
    \varphi_d' &= \dfrac{\sum_{t}K_{dt}}{\sum_x \sum_{\{rt;x\}} P_{dr} W_{x}} = \dfrac{\sum_{t}K_{dt}}{\sum_{r, t} P_{dr} W_{rt}} \; ,
    \label{eq:dragonfly_updateW}
\end{align}
where $\{rt;x\}$ indicates all pairs of $r$ and $t$ that rotate to a given voxel $x$.

In anticipation of the high-throughput XFEL-SPI experiments at LCLS-II, we simulated three one-million diffraction patterns using only elastic scattering \cite{Ayyer2016}. 
The consequential EMC reconstruction is both memory- and compute-intensive. 
Hence, the implementation of \eqref{eq:dragonfly_updateW} is efficiently distributed over GPUs (NVIDIA GTX 1080Ti) that are distributed over several compute nodes. 

When two latent structural models (labeled $m$) give rise to the diffraction patterns (Fig. \ref{fig:emc}a,d), the above equations could be 
modified by substituting $r$ with $m, r$:
\begin{align}
    W'_{mx} &= \dfrac{\sum_{d, m} \sum_{\{rt;x\}}P_{dmr}K_{dt}}{\sum_{d, m}
    \sum_{\{rt;x\}}P_{dmr}\varphi_d}\text{,} \\
        \varphi_d' &= \dfrac{\sum_{t}K_{dt}}{\sum_{m, x} \sum_{\{rt;x\}} P_{dmr} W_{x}} =
        \dfrac{\sum_{t}K_{dt}}{\sum_{m, r, t} P_{dr} W_{mrt}}\; .
\end{align}

\subsection{Conversion from phase differences to phase tilts}\label{sec:tilt}

For any given phase difference $\alpha$, i.e. between two neighboring pixels or the same pixel at different time points, it is possible to calculate a phase tilt angle
\begin{equation}
\theta = \frac{\alpha \lambda }{2 \pi \delta} 
\end{equation}
where $\lambda$ is the wavelength and $\delta$ the size of a pixel in the reconstructed wavefront. A more complete derivation for this simple relationship is given in Section 4 of \textcolor{blue}{Supplement 1}.
 
\medskip
\noindent\textbf{Funding.} filled automatically via Prism.
This project has received funding from the European Research Council (ERC) under the European Union's Seventh Framework Programme (Grant agreement No. 279753) and under the European Union's Horizon 2020 research and innovation programme (Grant agreement No. 866026). F.R.N.C.M acknowledges support from the Swedish Research Council (2018-00234). N.D.L recognizes the support of National University of Singapore (NUS) startup grant (R-154-000-A09-133), NUS Early career Research award (R-154-000-B35-133), and the Singapore National Research Foundation (NRF-CRP16-2015-05). 

\medskip
\noindent\textbf{Acknowledgments.} 
We thank Peter Walter for helpful discussions. We thank Björn Enders and Aaron Parsons for maintaining and developing the PtyPy software. Use of the Linac Coherent Light Source, SLAC National Accelerator Laboratory, is supported by the U.S. Department of Energy, Office of Science, Office of Basic Energy Sciences under Contract No. DE-AC02-76SF00515. We thank the LCLS staff for their assistance. We thank Bai Chang for maintaining the CBIS compute cluster.

\medskip
\noindent\textbf{Disclosures.} The authors declare no conflicts of interest.

\medskip
\noindent See \textcolor{blue}{Supplement 1} for supporting content. The ptychographic data sets used for the reconstructions (Fig. \ref{fig:recons}) and the orthogonalized probe modes (Fig. \ref{fig:beam}) are available on \textcolor{blue}{www.cxidb.org/id-178.html}. 

% Bibliography
\bibliography{references}

% Full bibliography added automatically for Optics Letters submissions; the following line will simply be ignored if submitting to other journals.
% Note that this extra page will not count against page length
\bibliographyfullrefs{references}

% Please include bios and photos of all authors for aop articles
\ifthenelse{\equal{\journalref}{aop}}{%
\section*{Author Biographies}
\begingroup
\setlength\intextsep{0pt}
\begin{minipage}[t][6.3cm][t]{1.0\textwidth} % Adjust height [6.3cm] as required for separation of bio photos.
  \begin{wrapfigure}{L}{0.25\textwidth}
    \includegraphics[width=0.25\textwidth]{john_smith.eps}
  \end{wrapfigure}
  \noindent
  {\bfseries John Smith} received his BSc (Mathematics) in 2000 from The University of Maryland. His research interests include lasers and optics.
\end{minipage}
\begin{minipage}{1.0\textwidth}
  \begin{wrapfigure}{L}{0.25\textwidth}
    \includegraphics[width=0.25\textwidth]{alice_smith.eps}
  \end{wrapfigure}
  \noindent
  {\bfseries Alice Smith} also received her BSc (Mathematics) in 2000 from The University of Maryland. Her research interests also include lasers and optics.
\end{minipage}
\endgroup
}{}

\end{document}